\documentclass[showpacs,amsmath,amssymb,aps,showkeys,floatfix,prd,a4paper]{revtex4}
\usepackage{subfigure}
\usepackage{dcolumn}
\usepackage{bm}
\usepackage{epsfig}
\usepackage{amsfonts}
\usepackage{amssymb,amscd}
\def\lsim{\raise0.3ex\hbox{$<$\kern-0.75em\raise-1.1ex\hbox{$\sim$}}}
\def\gsim{\raise0.3ex\hbox{$>$\kern-0.75em\raise-1.1ex\hbox{$\sim$}}}

\newcommand{\be}{\begin{equation}}
\newcommand{\ee}{\end{equation}}

\def\beq{\begin{equation}}
\def\eeq{\end{equation}}
\def\beqa{\begin{eqnarray}}
\def\eeqa{\end{eqnarray}}

\newcommand{\ba}{\begin{eqnarray}}

\def\gappeq{\mathrel{\rlap {\raise.5ex\hbox{$>$}}
{\lower.5ex\hbox{$\sim$}}}}
\def\lappeq{\mathrel{\rlap{\raise.5ex\hbox{$<$}}
{\lower.5ex\hbox{$\sim$}}}}
\def\Toprel#1\over#2{\mathrel{\mathop{#2}\limits^{#1}}}

\begin{document}
\begin{flushright}
\vskip1cm
\end{flushright}

\title{Production of exotic charmonium  in $\gamma \gamma$ interactions at hadronic colliders}
\author{B.D.  Moreira$^{1}$, C.A. Bertulani$^{2,3}$, V.P. Gon\c{c}alves$^{4}$  and  F.S. Navarra$^1$\\
$^1$Instituto de F\'{\i}sica, Universidade de S\~{a}o Paulo, C.P. 66318,  05315-970 S\~{a}o Paulo, SP, Brazil\\
$^2$Department of Physics and Astronomy, Texas A\&M University-Commerce, Commerce, Texas 75429, USA\\
$^3$Department of Physics and Astronomy, Texas A\&M University, College Station, Texas 77843, USA\\
$^{4}$High and Medium Energy Group, Instituto de F\'{\i}sica e Matem\'atica,  Universidade Federal de Pelotas\\
Caixa Postal 354,  96010-900, Pelotas, RS, Brazil.\\
}

\begin{abstract}
In this paper we investigate the Exotic Charmonium (EC) production in  $\gamma \gamma$ interactions present in proton-proton, proton-nucleus and nucleus-nucleus collisions at the CERN Large Hadron Collider (LHC) energies as well as for the proposed energies of the Future Circular Collider (FCC).  Our results demonstrate that the experimental study of these processes is feasible and can be used  to constrain the theoretical decay widths and shed some light on the configuration 
of the considered multiquark states.  
\end{abstract}

\pacs{12.38.-t, 24.85.+p, 25.30.-c}

\keywords{Quantum Chromodynamics, Exotic Vector Mesons, Photon -- photon interactions.}

\date{\today}

\maketitle


\section{Introduction} 

Over the last years the existence of exotic hadrons has been firmly established \cite{hosa,espo,nnl} and now the next step is to accurately determine their 
structure.  Among the proposed configurations, the  meson molecule  and the tetraquark are the most often discussed. 
The main difference between a tetraquark and  a meson molecule  is that the former is compact and the interaction between the constituents occurs  through 
color exchange  forces whereas the latter is an extended object and the interaction between its constituents  happens through meson exchange forces.  It is also 
possible that the observed states are   charmonium-tetraquark,  charmonium-molecule or tetraquark-molecule mixtures. Indeed this mixed approach has led to the best 
description of the $X(3872)$. In Ref. \cite{zanetti} the mass and strong decay width were very well reproduced assuming that the $X(3872)$ has a $c \bar{c}$ 
component with a weight of  97 \% and a $D \bar{D}^*$ component with  $3$ \% weight. As for the production in proton-proton (pp) collisions, both at Fermilab 
and at the LHC, in Ref. \cite{chines} it was shown that the best description can be achieved with a charmonium-molecule combination, i.e. $\chi_{c1}' - D \bar{D}^*$, 
in which the $c \bar{c}$ component is of the order of $28 - 44$ \%.  Even if the best description is given by a mixture it is still very important to understand the 
individual role played by each component.

One of the reactions which were proposed as a tool to discriminate between the two theoretical descriptions of the exotic states ($R$) is the decay into two photons, 
i.e., $R \rightarrow \gamma \gamma$.  This process involves  particle-antiparticle annihilation, which is sensitive to the spatial configuration of the decaying 
states and should be hindered  if its constituents are away from each other, as it is the case in a molecular configuration.  In fact, for an S-wave non-relativistic 
two-body  system $R$ in a state described by a wave function $\psi(r)$ the width for annihilation into $\gamma \gamma$ is given by
\beq
\Gamma (R \rightarrow \gamma \gamma ) = \frac{2 \pi \alpha^2}{M^2_R} \, |\psi(0)|^2
\label{largura}
\eeq
We may expect that for a loosely bound meson molecule $|\psi(0)|^2$ is much smaller than for a diquark-antidiquark compact system.

The production of exotic particles 
in hadronic colliders is one  of the most promising testing  grounds for our ideas about the structure of the new states. It has  been shown 
\cite{espo,espo2,espo3} that it is  difficult to produce molecules in pp  collisions. In a pure  molecular approach the estimated cross section for 
$X(3872)$ production 
is two orders of magnitude smaller than the measured one. One  might try to  explain these data with a pure tetraquark model. An attempt to do this, using an 
extension of 
the  color evaporation model to the cases where we have double parton scattering, was presented in \cite{tetradps}. An alternative is to explore the 
fact that ultra-relativistic hadrons are an intense source of photons (For a review see
Ref. \cite{BB88,BB94,BKN05,Bal08,hencken,Baur02}) and investigate  resonance production in the $\gamma \gamma$ and $\gamma h$ ($h$ = p, A) interactions 
present in pp/pA/AA collisions. At large impact parameters ($b > R_{h_1} + R_{h_2}$), denoted hereafter ultra - peripheral collisions (UPCs), the photon -- 
induced interactions become dominant with the final state  being  characterized by the state $R$ and  the presence of one intact hadron, in the case of an 
inclusive $\gamma h$ interaction, or two intact hadrons if the resonance was produced in a $\gamma \gamma$ or a diffractive $\gamma h$ interactions. 
Recent experimental results at
RHIC \cite{star,phenix}, Tevatron \cite{cdf} and LHC \cite{alice, alice2,lhcb,lhcb2,lhcb_ups,cms1,cms2,cms3,Atlas}  have demonstrated that the study of 
photon -- induced interactions in hadronic collisions is feasible and can be used to improve our understanding of the QCD dynamics as well to probe Beyond 
Standard Model Physics (See e.g. Refs. \cite{vicber,bruno,vicwersm}). In this work we will systematically explore the possibility of producing Exotic 
Charmonium (EC)  in two-photon interactions in UPCs  with  
ultra-relativistic protons and nuclei. We consider hadronic collisions at the LHC as well as in the proposed Future Circular Collider (FCC) \cite{fcc}.    

The idea of studying exotic meson production in UPCs was pioneered in \cite{bertu}, where the production cross section of several light and heavy well 
known mesons (and also  exotic mesons and glueballs candidates) in nucleus-nucleus collisions was computed. 
Later, in Ref. \cite{vicwer}, the same formalism was applied to the production of mesons and exotic states in proton-proton collisions. Special attention 
was given  to the exotic charmonium states $X(3940)$ and $X(4140)$. More recently, in Ref. \cite{vicmar}, the authors calculated the cross sections of the processes 
pp $\rightarrow$ pn$X$, where $X$ are the  exotic charmonium states $Z_c (3900)$, $Z(4430)$, $X(3940)$ and $X(3915)$. In these reaction one proton emits 
one  photon and the other emits a  pion or a Pomeron.

In this work we revisit and update the calculations performed in \cite{bertu} and \cite{vicwer}, extending them to  pp, pA and AA  collisions at LHC and FCC energies. We shall focus on photon-photon production of the exotic charmonium states and include $X(3915)$, $Z(3930)$ and $X(4160)$. 
As it will be seen, all the ingredients of the calculation are fixed with the exception of the two-photon decay width of the exotic state (\ref{largura}).
In principle tetraquark 
and molecular configurations would yield quite different numbers for the decay widths, which would yield quite different production cross sections. The two-photon 
decay width of the exotic states has been calculated in the molecular approach in several works \cite{branz1,branz2,branz3,volo}.  Unfortunately, the
theoretical predictions of the tetraquark model are not yet available. 
We are going to present 
production cross sections of meson molecules keeping in mind that, if the states in question were tetraquarks, the corresponding cross sections would be 
much larger.

This paper is organized as follows. In section II we present  a short description of the formalism used for particle production in $\gamma \gamma$ interactions at 
hadronic colliders. In section III we present our predictions for the exotic charmonium production in pp/pA/AA collisions at LHC and FCC energies. Finally, in 
section IV we summarize our main conclusions.

\section{Formalism}

\begin{figure}[t]
\centering
\includegraphics[scale=0.5]{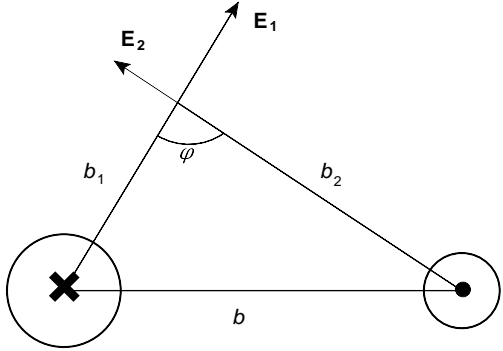}
\caption{Electromagnetic field interaction in ultraperipheral hadron-hadron (or nucleus-nucleus) collisions. The particle on the left moves into the page and 
the particle on the right moves out of the page. They are separated by the impact parameter b.}
\label{esquema_colisao}
\end{figure}

Since the theoretical treatment of UPCs in relativistic heavy ion collisions has been extensively discussed  in the literature 
\cite{BB88,BB94,BKN05,Bal08,hencken,Baur02}, in what follows we will only review the main formulas needed to make predictions for exotic meson production 
in $\gamma \gamma$ interactions.    In the equivalent photon approximation, the cross section for the production of a generic exotic charmonium 
state, $X$,    in UPCs between two hadrons, $h_{1}$ and $h_{2}$, is given by (See e.g. \cite{BB88,hencken})
\begin{eqnarray}
\sigma \left( h_1 h_2 \rightarrow h_1 \otimes R \otimes h_2 ;s \right)   
&=& \int \hat{\sigma}\left(\gamma \gamma \rightarrow R ; 
W \right )  N\left(\omega_{1},{\mathbf b_{1}}  \right )
 N\left(\omega_{2},{\mathbf b_{2}}  \right ) S^2_{abs}({\mathbf b})  
 \mbox{d}^{2} {\mathbf b_{1}}
\mbox{d}^{2} {\mathbf b_{2}} 
\mbox{d} \omega_{1}
\mbox{d} \omega_{2} \,\,\, ,
\label{sec_hh}
\end{eqnarray}
where $\sqrt{s}$ is center-of-mass energy for the $h_1 h_2$ collision ($h_i$ = p,A), $\otimes$ characterizes a rapidity gap in the final state 
and $W = \sqrt{4 \omega_1 \omega_2}$ is the invariant mass of the $\gamma \gamma$ system. Moreover, 
 $N(\omega_i,b_i)$ is the equivalent photon spectrum generated by hadron (nucleus) $i$, and $\sigma_{\gamma \gamma \rightarrow R}(\omega_{1},\omega_{2})$ 
is the cross section for the production of a state $R$ from two real photons with energies $\omega_1$ and $\omega_2$. Moreover, in Eq.(\ref{sec_hh}),
$\omega_{i}$ is the energy of the photon emitted by the hadron (nucleus) $h_{i}$ at an impact parameter, or distance, $b_{i}$ from $h_i$. The photons, and 
their  corresponding electric fields, interact at the point  shown in Fig. \ref{esquema_colisao}.
The factor $S^2_{abs}({\mathbf b})$ is the absorption factor, given in what follows by \cite{BF90}
\begin{eqnarray}
S^2_{abs}({\mathbf b}) = \Theta\left(
\left|{\mathbf b}\right| - R_{h_1} - R_{h_2}
 \right )  = 
\Theta\left(
\left|{\mathbf b_{1}} - {\mathbf b_{2}}  \right| - R_{h_1} - R_{h_2}
 \right )  \,\,,
\label{abs}
\end{eqnarray}
where $R_{h_i}$ is the radius of the hadron $h_i$ ($i = 1,2$). 
The presence of this factor in Eq. (\ref{sec_hh})  excludes the overlap between the colliding hadrons and allows to take into account only ultraperipheral collisions.
Remembering  that the photon energies $\omega_1$ and $\omega_2$  are related to   
$W$ and the rapidity $Y $ of the outgoing resonance $R$ by 
\begin{eqnarray}
\omega_1 = \frac{W}{2} e^Y \,\,\,\,\mbox{and}\,\,\,\,\omega_2 = \frac{W}{2} e^{-Y} \,\,\,
\label{ome}
\end{eqnarray}
the total cross section can be expressed by (For details see e.g. Ref. \cite{kluga})
\begin{eqnarray}
\sigma \left( h_1 h_2 \rightarrow h_1 \otimes R \otimes h_2 ;s \right)   
&=& \int \hat{\sigma}\left(\gamma \gamma \rightarrow R ; 
W \right )  N\left(\omega_{1},{\mathbf b_{1}}  \right )
 N\left(\omega_{2},{\mathbf b_{2}}  \right ) S^2_{abs}({\mathbf b})  
\frac{W}{2} \mbox{d}^{2} {\mathbf b_{1}}
\mbox{d}^{2} {\mathbf b_{2}} 
\mbox{d}W 
\mbox{d}Y \,\,\, .
\label{cross-sec-2}
\end{eqnarray}
The equivalent photon flux can be expressed  as follows 
\begin{equation}
N(\omega,b) = \frac{Z^{2}\alpha_{em}}{\pi^2}\frac{1}{b^{2}\omega}
\left[ \int u^{2} J_{1}(u) F\left(\sqrt{\frac{\left( {b\omega}/{\gamma}\right)^{2} + u^{2}}{b^{2}}} \right )
\frac{1}{\left({b\omega}/{\gamma}\right)^{2} + u^{2}} \mbox{d}u\right]^{2}\, ,
\label{fluxo}
\end{equation}
where $F$ is the nuclear form factor of the  equivalent photon source. 
In the nuclear case, it is often used in the literature  a monopole form factor given by \cite{kluga}
\begin{equation}
F(q) = \frac{\Lambda^{2}}{\Lambda^{2} + q^{2}} \, ,
\label{ff_nuc}
\end{equation}
with $\Lambda = 0.088$ GeV. For proton projectiles, the form factor is in general assumed to be \cite{Goncalves:2015sfy,Goncalves:2016ybl}
\begin{eqnarray}
F(q) = 1/
\left[1 + q^{2}/(0.71\mbox{GeV}^{2}) \right ]^{2} \, .
\label{ff_pro}
\end{eqnarray}
In what follows we will assume these form factors  to estimate the cross sections. However, as discussed in detail in Ref. \cite{kluga}, distinct models for $F$ imply that the resulting cross sections can differ significantly. In order to estimate the theoretical uncertainty associated to the model used for $F$, in what follows we also will present the predictions obtained assuming  $ F(q) = 1$, i.e.  that proton and nucleus are point-like particles. In this case, we need to integrate  
from a minimum distance $b_i=R_i$ ($\sim 0.7$ fm for protons and $1.2A^{1/3}$ fm for nuclei) in Eq. \eqref{sec_hh}, because the flux is divergent 
for $b=0$ \cite{bertu,JAC}. Additionally,  in the case of $PbPb$ collisions, we also will consider a more realistic form factor, obtained as a Fourier transform of the Woods - Saxon distribution for the nuclear density. As demonstrated in Ref.  \cite{kluga}, this form factor coincides with the monopole one only in a very limited range of values of the photon virtuality, with the difference between them becoming larger at large values of $q$.

In order to estimate the $ h_1 h_2 \rightarrow h_1 \otimes R \otimes h_2$ cross section we need the $\gamma \gamma \rightarrow R$ interaction cross section as input. 
In what follows  we will use the Low formula \cite{Low}, where the cross section for the production of  the $R$ 
state due to the two-photon fusion can be written in terms of the two-photon decay width of  the corresponding state as  
\begin{eqnarray}
 \sigma_{\gamma \gamma \rightarrow R}(\omega_{1},\omega_{2}) = 
8\pi^{2} (2J+1) \frac{\Gamma_{R \rightarrow \gamma \gamma}}{M_{R}} 
\delta(4\omega_{1}\omega_{2} - M_{R}^{2}) \, ,
\label{Low_cs}
\end{eqnarray}
where the decay width $\Gamma_{R \rightarrow \gamma \gamma}$ can in some cases be taken from experiment or can be theoretically estimated. 
Furthermore, $M_{R}$ and $J$ are, respectively, the mass and spin of the  produced  state.
Finally, it is important to emphasize that due to the $Z^2$ dependence of the photon spectra, we have that for the same $W$ the following hierarchy is expected 
to be valid for the resonance production induced by $\gamma \gamma$ interactions: $\sigma_{AA} = Z^2 \cdot \sigma_{pA} = Z^4 \cdot \sigma_{pp}$.

\section{Results}

\begin{figure}
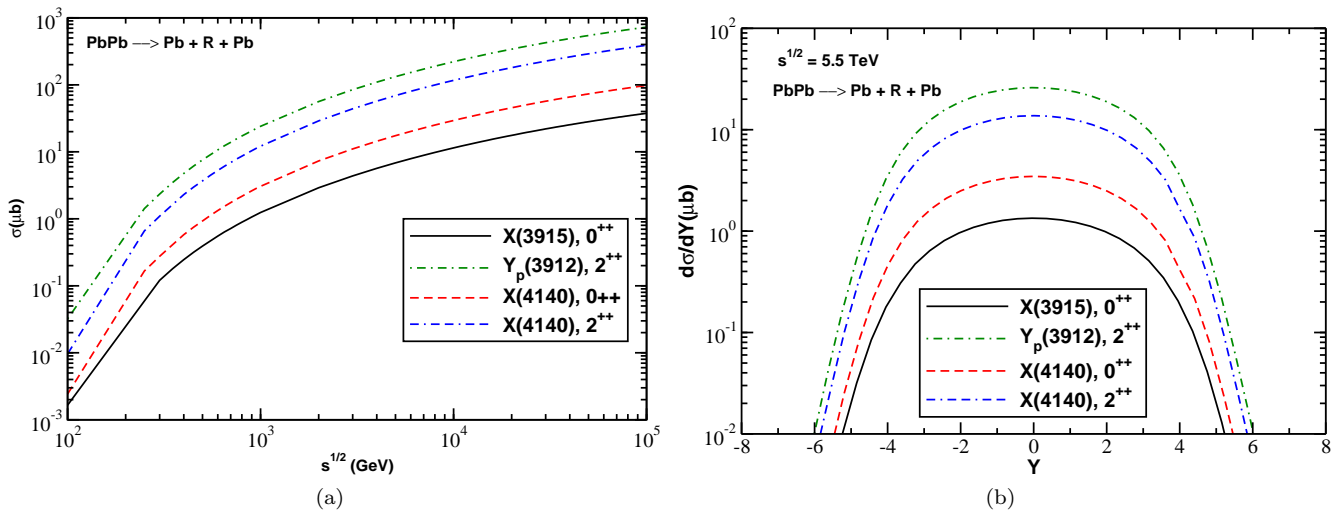

\begin{center}
\subfigure[ ]{\label{figa}
\includegraphics[width=0.48\textwidth]{NOVO_energia_AA.eps}}
\subfigure[ ]{\label{figb}
\includegraphics[width=0.48\textwidth]{NOVO_rapidez_AA.eps}}
\end{center}
\caption{(a) Cross section  of the process PbPb $\rightarrow$ Pb$\otimes$R$\otimes$Pb  as a function of the energy $\sqrt{s}$. (b) Rapidity distribution 
of the  resonance produced in  $\gamma \gamma$ interactions in Pb-Pb collisions at $\sqrt{s} = 5.5$ TeV. 
}
\label{fig2}
\end{figure}

In this Section we present our predictions for the production of exotic mesons due to photon-photon fusion in UPCs at energies available at the LHC and proposed 
for the FCC.  We have considered all 
the charmonium states for which either a measurement or a theoretical estimate of the decay width is available. For the sake of comparison with the results found in 
\cite{vicmar} we consider the two possible assignments, $0^{++}$ and $2^{++}$,  for the states $X(3940)$ and $X(4140)$. In fact, in the last edition of the PDG
\cite{pdg} these states still appear with undefined assignments. 
The masses and decay widths were inferred from Refs. \cite{branz1,branz2,branz3}. We use the following notation: $\sigma_{b_{min}}$ denotes cross sections 
evaluated with $F = 1$ and $\sigma_{F}$ denotes cross sections evaluated with the form factors from Eqs. (\ref{ff_nuc}) and (\ref{ff_pro}) for nuclei and protons, 
respectively. In the particular case of $PbPb$ collisions we also will present the predictions obtained using the realistic form factor \cite{kluga}, which we will denote by $\sigma_{R}$.  The precise form of the  form factor is the main source of uncertainties in our calculations and the use of the two cases mentioned above gives us an estimate of the theoretical error. 

Initially let us consider the energy dependence of the total cross sections and the rapidity distributions of the resonances  produced in 
$\gamma \gamma$ interactions in UPCs. These observables were shown to be the most useful ones to be  compared with theoretical predictions. 
This expectation has been confirmed by recent experimental results (obtained at  RHIC  and also at the LHC) on vector meson production 
($\rho$, J/$\Psi$ and $\Upsilon$) \cite{star,phenix,cdf,alice, alice2,lhcb,lhcb2,lhcb_ups,cms1,cms2,cms3,Atlas}. Here we propose to extend these measurements 
beyond the production of well-established mesons, such as the J/$\Psi$, and use UPCs in hadronic colliders to asses new  information on 
exotic mesons and constrain theoretical predictions. 
In Fig. \ref{fig2}a, we present our predictions for the energy dependence  of the production cross section in Pb-Pb collisions with $\sqrt{s}$ from 
100 GeV to 100 TeV obtained using the monopole form factors and the widths presented in Table \ref{sig_AA}. Similar energy dependences are predicted for 
p-Pb and p-p collisions, with the normalization scaled by a factor $\approx 1/Z^2$ and $\approx 1/Z^4$, respectively. The predicted cross 
sections for the LHC kinematical range are of the order of 1 -- 100 $\mu b$. Moreover, this result shows us that the cross sections are one order of 
magnitude larger for the energies expected to be covered by the FCC in Pb-Pb collisions ($\sqrt{s} = 39$ TeV). 
In Fig. \ref{fig2}b, we show the rapidity distribution of the exotic charmonium production  in Pb-Pb collisions at $\sqrt{s} = 5.5$ TeV. We have that 
the maximum of the distribution occurs at central rapidities, strongly decreasing at forward and backward rapidities. In particular, for the $X(4140)$ 
production, the two predictions differ by a factor 3 at $Y = 0$. 

\begin{table}[t]
\begin{center}
\begin{tabular}{|c|c|c|c|c|c|c|c|c|c|c|c|}
\hline 
State & Mass & $\Gamma_{\gamma\gamma}^{theor}$(keV) & \multicolumn{3}{c|}{$\sigma_{b_{min}}$ ($\mu$b)} & \multicolumn{3}{c|}{$\sigma_{F}$ ($\mu$b)}& \multicolumn{3}{c|}{$\sigma_{R}$ ($\mu$b)}\tabularnewline
\cline{4-12} 
 &  & &  $2.76$ TeV & $5.5$ TeV & $39$ TeV  & $2.76$ TeV & $5.5$ TeV & $39$ TeV & $2.76$ TeV & $5.5$ TeV & $39$ TeV \tabularnewline
    \hline
    \hline
    X(3940), 0$^{++}$ & 3943 &  0.33  &  4.2 &   8.2  & 31.6 &   6.5 &  11.8  & 40.9   & 5.7 & 10.8 & 39.6 \tabularnewline  
    X(3940), 2$^{++}$ & 3943 & 0.27  & 17.2  & 33.6 & 129.2 & 26.5 &   48.4   & 167.4   & 23.4 & 44.2 & 162.0 \tabularnewline
    X(4140), 0$^{++}$ & 4143 & 0.63  & 6.5 &  12.9  & 51.2  & 10.2 &  18.7  & 65.7 & 9.0 & 17.1 & 63.6         \tabularnewline
    X(4140), 2$^{++}$ & 4143 & 0.50  & 26.0 &  51.2 & 201.0  & 40.3 &  74.3   & 260.6      & 35.5 & 67.7 & 252.3 \tabularnewline
    Z(3930), 2$^{++}$ & 3922 &      0.083       & 5.4  & 10.5  & 40.9  & 8.3 & 15.2  & 52.4    & 7.4 & 13.9 & 50.5   \tabularnewline
    X(4160), 2$^{++}$ & 4169 &      0.363       & 18.4 &   36.4 & 144.2  &28.6 & 52.7  & 185.3    & 25.2 & 48.1 & 178.7  \tabularnewline
    Y$_{p}$(3912), 2$^{++}$ & 3919 &   0.774    & 50.5 &   98.6 & 382.4 &  77.9 &  142.2  & 490.1   & 68.9 & 129.9 & 473.7  \tabularnewline
    X(3915), 0$^{++}$ & 3919 &   0.20    & 2.6 &   5.1 & 19.8 &  4.0 &  7.3  & 25.3 & 3.6 & 6.7 & 24.5  \tabularnewline
    \hline
\end{tabular}
\caption{Cross sections for exotic meson production in Pb-Pb collisions using the theoretical decay rates presented in Refs.  \cite{branz1,branz2,branz3}.}
\label{sig_AA}
\end{center}
\end{table}

In Tables \ref{sig_AA}, \ref{sig_pA} and \ref{sig_pp} we present our predictions for the exotic charmonium production in Pb-Pb, p-Pb and p-p collisions, 
respectively, using the form factors mentioned in the previous Section. 
Owing to the form of the cross section of Eq. (\ref{sec_hh}) and its dependence on the equivalent photon spectrum (\ref{fluxo}), the Pb-Pb cross 
sections are enhanced  by a factor $Z^{4} \, (Z^2)$ in comparison to p-p (p-Pb) collisions. This is reflected in our calculations, with the cross 
sections ranging from a few hundred nb up to  hundred of $\mu$b. 

In Table \ref{sig_AA} we present our predictions for the cross sections for the production of several exotic mesons in Pb-Pb collisions at  
$\sqrt{s} = 2.76$ TeV, $\sqrt{s} = 5.5$ TeV and $\sqrt{s} = 39$ TeV.   Comparing the cross sections for different form factors we observe that 
$\sigma_{F} \, \approx \, 1.5 \, \sigma_{b_{min}}$. This happens because $\sigma_{b_{min}}$  does not take into account  meson production in the region  
$b_{i} < R_{i}$, while $\sigma_{F}$ allows for this, as long as the constraint 
$b>R_{1}+R_{2}$ is respected. Since the masses of the exotic states are nearly the same (within 5\%), the main sources of changes in the cross sections are  the 
magnitude of the decay width and the spin of the produced particle. The predicted cross sections are of the order of $\mu b$ and increase with the energy, as 
expected from Fig. \ref{fig2}.  We can see that the predictions for the $X(3940)$ differ by a factor 4, depending on the spin assumed for the particle. Similar 
differences are predicted in the case of  $X(4140)$ production. An important aspect is that the predictions for the production of the $X(3915)$ and $Y_p(3912)$ differ 
by a factor 20. Currently, it is not clear if these states are the same or not. Consequently, our results indicate that the study of their production in UPCs 
can be useful to constrain their main characteristics. 

\begin{table}[t]
\begin{center}
\begin{tabular}{|c|c|c|c|c|c|c||c|c|}
\hline 
State & Mass & $\Gamma_{\gamma\gamma}^{theor}$(keV) & \multicolumn{3}{c|}{$\sigma_{b_{min}}$ (nb)} & \multicolumn{3}{c|}{$\sigma_{F}$ (nb)}\tabularnewline
\cline{4-9} 
 &  & &  $5$ TeV & $8.8$ TeV & $63$ TeV & $5$ TeV & $8.8$ TeV & $63$ TeV \tabularnewline
    \hline
    \hline
    X(3940), 0$^{++}$ & 3943 &  0.33  &  2.8 &  4.0 & 10.6 &   3.3 &  4.5  & 11.3     \tabularnewline  
    X(3940), 2$^{++}$ & 3943 & 0.27  & 11.4  & 16.3 & 43.4 & 12.9 &   18.3  & 46.3     \tabularnewline
    X(4140), 0$^{++}$ & 4143 & 0.63  & 4.4 &  6.3  & 16.6 & 5.0 &  7.1  & 18.3      \tabularnewline
    X(4140), 2$^{++}$ & 4143 & 0.50  & 17.6 &  25.2  & 65.9 & 20.0 &  28.4  & 72.5       \tabularnewline
    Z(3930), 2$^{++}$ & 3922 &      0.083       & 3.6  & 5.1  & 13.2  & 4.0 & 5.7  & 14.5     \tabularnewline
    X(4160), 2$^{++}$ & 4169 &      0.363       & 12.5 &   17.9 & 46.9 & 14.2 & 20.1  & 63.3     \tabularnewline
    Y$_{p}$(3912), 2$^{++}$ & 3919 &   0.774    & 33.5 &   47.7 & 123.3 & 37.9 &  53.6 & 132.0     \tabularnewline
    X(3915), 0$^{++}$ & 3919 &   0.20    & 1.7 &   2.5 & 6.4 & 2.0 &  2.8  & 7.0    \tabularnewline
    \hline
\end{tabular}
\caption{Cross sections for exotic meson production in p-Pb collisions using the  theoretical decay rates presented in Refs. \cite{branz1,branz2,branz3}.}
\label{sig_pA}
\end{center}
\end{table}

In Table \ref{sig_pA}, we present our results for the production of exotic mesons in p-Pb collisions at $\sqrt{s} = 5$ TeV, $8.8$ TeV and $63$ TeV.  In this case 
we can observe that the differences between the predictions obtained with $\sigma_{b_{min}}$ and $\sigma_{F}$ are smaller than in the Pb-Pb case. This occurs 
because  the effects of meson production in the region $b_{i}<R_{i}$, calculated with  Eq. (\ref{ff_pro}), are attenuated by the fact that the proton has a smaller radius 
than the Pb.  Furthermore, in this case, the cross section is enhanced by a factor $Z^{2}$ in comparison to the p-p one, leading to cross sections that can 
only reach a few tens of nb. The differences between the different predictions, observed in the $A - A$ case, also are present in p-Pb collisions.

\begin{table}[t]
\begin{center}
\begin{tabular}{|c|c|c|c|c|c|c|c|c|}
\hline 
State & Mass & $\Gamma_{\gamma\gamma}^{theor}$(keV) & \multicolumn{3}{c|}{$\sigma_{b_{min}}$ (pb)} & \multicolumn{3}{c|}{$\sigma_{F}$ (pb)}\tabularnewline
\cline{4-9} 
 &  & &  $7$ TeV & $14$ TeV & $100$ TeV & $7$ TeV & $14$ TeV & $100$ TeV \tabularnewline
    \hline
    \hline
    X(3940), 0$^{++}$ & 3943 &  0.33  &  0.98 &  1.3 & 2.8 &   1.0 &  1.5  & 2.8     \tabularnewline  
    X(3940), 2$^{++}$ & 3943 & 0.27  & 4.0  & 5.6 & 11.4 & 4.1 &   5.7  & 11.6     \tabularnewline
    X(4140), 0$^{++}$ & 4143 & 0.63  & 1.6 &  2.2 & 4.5  & 1.6 &  2.2  & 4.6        \tabularnewline
    X(4140), 2$^{++}$ & 4143 & 0.50  & 6.2 &  8.7  & 18.0 & 6.4 &  8.9  & 18.3       \tabularnewline
    Z(3930), 2$^{++}$ & 3922 &      0.083       & 1.2  & 1.7  & 3.6  & 1.3 & 1.8  & 3.6     \tabularnewline
    X(4160), 2$^{++}$ & 4169 &      0.363       & 4.4 &   6.1 & 12.8 & 4.5 & 6.3  & 13.0     \tabularnewline
    Y$_{p}$(3912), 2$^{++}$ & 3919 &   0.774    & 11.7 &   16.3 & 33.4 & 12.0 &  16.7  & 34.0    \tabularnewline
    X(3915), 0$^{++}$ & 3919 &   0.20    & 0.60 &   0.84 & 1.7 & 0.62 &  0.86  & 1.8    \tabularnewline
    \hline
\end{tabular}
\caption{Cross sections for exotic meson production in pp collisions using the theoretical decay rates presented in Refs.  \cite{branz1,branz2,branz3}.}
\label{sig_pp}
\end{center}
\end{table}

In Table \ref{sig_pp} we present our results for the production of exotic mesons  in p-p collisions at $\sqrt{s} = 7$ TeV, $14$ TeV and 100 TeV.  Here we observe a 
smaller difference between the two choices of form factor when compared with the previous cases. Moreover, in this case ($Z=1$) we do not have any  
enhancement of the cross section compared with the other cases, leading to much smaller cross sections. Even so, these are non-negligible cross sections, of 
order of few pb, well within reach of present experiment detection techniques, considering the high luminosity present in pp collisions. 

Before concluding, let us compare our predictions for the production of the  $X(3915)$ and $X(3940)$ states in $\gamma \gamma$ interactions with those 
obtained in Ref. \cite{vicmar}, where the contribution associated to $\gamma h$ interactions in pp collisions was estimated. We observe that the cross 
sections obtained in Ref. \cite{vicmar} are of the order of $nb$, while our predictions, presented in Table \ref{sig_pp}, are of the order of $pb$. 
Therefore, the dominant channel for the production of these states are  $\gamma h$ interactions. However, as demonstrated in Ref. \cite{vicmar}, they  
will be produced in the very forward region, with a large background associated to the Pomeron exchange, which makes the experimental separation of these 
states a hard task. In contrast, in $\gamma \gamma$ interactions, they will produced essentially at central rapidities as shown in Fig. \ref{fig2}(b), 
i.e. in the kinematical range covered by the current LHC detectors.

\section{Conclusions}

In this work we have studied the production of exotic mesons in UPCs at LHC and FCC energies due to two photon fusion. This is a clean process where the particles of the 
initial state are intact at the final state and can be detected at the forward direction as featured by the presence of two rapidity gaps between the 
projectiles and the produced particle. Moreover, we have predicted large values for the cross sections in PbPb and pPb collisions and non-negligible values in 
pp collisions. Our predictions for the rapidity distributions can also be of relevance for testing the theoretical models used in the calculations.  Therefore, 
we conclude that the experimental study is worth pursuing, that it can be useful to constrain decay widths evaluated theoretically and, ultimately, it can help 
in determining the  configuration of the considered multiquark states.

\begin{acknowledgments}
This work was  partially financed by the Brazilian funding agencies CNPq, CAPES, FAPERGS and FAPESP. This work was supported in part by the U.S. DOE 
grant DE- FG02-08ER41533 and the U.S. NSF Grant No. 1415656.
\end{acknowledgments}

\hspace{1.0cm}

\end{document}